\documentclass[11pt]{article}
\usepackage{fullpage,graphicx}
\usepackage[normalem]{ulem}

\newtheorem{theorem}{Theorem}
\newtheorem{lemma}[theorem]{Lemma}

\newcommand{\SA}{\ensuremath{\mathrm{SA}}}
\newcommand{\BWT}{\ensuremath{\mathrm{BWT}}}
\newcommand{\LCP}{\ensuremath{\mathrm{LCP}}}
\newcommand{\PLCP}{\ensuremath{\mathrm{PLCP}}}
\newcommand{\LF}{\ensuremath{\mathrm{LF}}}
\newcommand{\LCE}{\ensuremath{\mathrm{LCE}}}
\newcommand{\MS}{\ensuremath{\mathrm{MS}}}
\newcommand{\head}{\ensuremath{\mathrm{head}}}
\newcommand{\tail}{\ensuremath{\mathrm{tail}}}
\newcommand{\finger}{\ensuremath{\mathrm{finger}}}
\newcommand{\pos}{\ensuremath{\mathrm{pos}}}
\newcommand{\len}{\ensuremath{\mathrm{len}}}
\newcommand{\offset}{\ensuremath{\mathrm{offset}}}

\begin{document}

\title{MONI can find $k$-MEMs}
\author{Igor Tatarnikov, Ardavan Shahrabi Farahani,\\
Sana Kashgouli and Travis Gagie}
\maketitle

\begin{abstract}
Suppose we are asked to index a text $T [0..n - 1]$ such that, given a pattern $P [0..m - 1]$, we can quickly report the maximal substrings of $P$ that each occur in $T$ at least $k$ times.  We first show how we can add $O (r \log n)$ bits to Rossi et al.'s recent MONI index, where $r$ is the number of runs in the Burrows-Wheeler Transform of $T$, such that it supports such queries in $O (k m \log n)$ time.  We then show how, if we are given $k$ at construction time, we can reduce the query time to $O (m \log n)$.
\end{abstract}

\section{Introduction}

In his foundational text {\it Compact Data Structures: A Practical Approach}~\cite[Section 11.6.1]{Nav16}, Navarro posed the following problem:
\begin{quotation}
``Assume we have the suffix tree of a collection of genomes $T [0..n - 1]$.  We then receive a short DNA sequence $P [0..m - 1]$ and want to output all the maximal substrings of $P$ that appear at least $k$ times in $T$\ldots  Those substrings of $P$ are likely to have biological significance.''\footnote{We have changed $T$ to $T [0..n - 1]$ and $P [1..m]$ to $P [0..m - 1]$ for consistency with the rest of this paper, and omitted a parameter bounding from below the length of the substrings (since we can filter them afterwards).}
\end{quotation}
He described how to solve the problem with a suffix tree for $T$ in $O (m\,\mathrm{polylog} (n))$ time.  Since $T$ is a collection of genomes, it is likely to be highly repetitive and the theoretically best suffix-tree implementation is likely to be the $O (r \log (n / r))$-space one by Gagie, Navarro and Prezza~\cite{GNP20}, where $r$ is the number of runs in the Burrows-Wheeler Transform (BWT) of $T$.

Very recently, Navarro~\cite{Nav22} also gave solutions not based on a suffix tree, with the following bounds:
\begin{itemize}
\item $O (g_\mathit{rl})$ space and $O (k m^2 \log^\epsilon n)$ query time;
\item $O (\delta \log (n / \delta))$ space and $O (m \log m (\log m + k \log^\epsilon n))$ query time;
\item $O (g)$ space and $O (m^2 \log^{2 + \epsilon} n)$ query time when $k = \omega (\log^2 n)$;
\item $O (\gamma \log (n / \gamma))$ space and $O (m \log m \log^{2 + \epsilon} n)$ query time when $k = \omega (\log^2 n)$.
\end{itemize}
We refer readers to Navarro's paper and the references therein for definitions of $g_\mathit{rl}$, $\delta$, $g$ and $\gamma$.

In this paper we first show how we can add $O (r \log n)$ bits to Rossi et al.'s~\cite{ROLGB22} recent MONI index to obtain a solution with $O (k m \log n)$ query time.  We then show how, if we are given $k$ at construction time, we can reduce the query time to $O (m \log n)$.  The paper is laid out as follows: in Section~\ref{sec:MONI} we review MONI in enough depth to build on it; in Section~\ref{sec:LCP} we show how we can extend $\phi$ queries to support sequential access to the LCP array; in Section~\ref{sec:slow} we show how to use $\phi$ and $\phi^{-1}$ queries and LCP access to obtain a solution with $O (k m \log n)$ query time; we conclude in Section~\ref{sec:fast} by showing how, if we are given $k$ at construction time, we can pre-compute some answers, reducing the query time to $O (m \log n)$.  For the sake of brevity we assume readers are familiar with the concepts in Navarro's text.

\section{MONI}
\label{sec:MONI}

Bannai, Gagie and I~\cite{BGI20} designed an index for $T$ that takes $O (r \log n)$ bits plus the space needed to support fast random access to $T$, and lists all the maximal exact matches (MEMs) of $P$ with respect to $T$ --- that is, all the substrings $P [i..j]$ of $P$ occurring in $T$ such that $i = 0$ or $P [i - 1..j]$ does not occur in $T$, or $j = m - 1$ or $P [i..j + 1]$ does not occur in $T$ --- in $O (m \log \log n)$ time plus the time needed for $O (m)$ random accesses to $T$.  MEMs are widely used in DNA alignment~\cite{Li13} and they are the substrings of $P$ Navarro asks for when $k = 1$.  Generalizing to arbitrary $k$, we refer to the substrings he asks for as $k$-MEMs.

Bannai et al.\ did not give an efficient construction algorithm or an implementation, but Rossi et al.\ later did.  They called their implementation MONI, the Finnish word for ``multi'', since it is intended to store a multi-genome reference.  Boucher et al.~\cite{BGIKLMNPR21} then gave a version of MONI that processes $P$ online using longest common extension (LCE) queries on $T$ instead of random access.  We can support those LCE queries in $O (\log n)$ time with a balanced straight-line program for $T$, which in practice takes significantly less space than the rest of MONI.

We now sketch how Boucher et al.'s version of MONI works, incorporating ideas from Nishimoto and Tabei~\cite{NT21} and Brown, Gagie and Rossi~\cite{BGR22} about replacing rank queries by table lookup and assuming we have an LCE data structure.  Suppose
\[T = \mathtt{GATTACAT\#AGATACAT\#GATACAT\#GATTAGAT\#GATTAGATA\$}\]
with $\mathtt{\$} \prec \mathtt{\#} \prec \mathtt{A} \prec \cdots \prec \mathtt{T}$, and consider Table~\ref{tab:base_table}, in which the permutation FL is just the inverse of the more familiar permutation LF.

\begin{table}
\begin{center}
\includegraphics[height=.9\textheight]{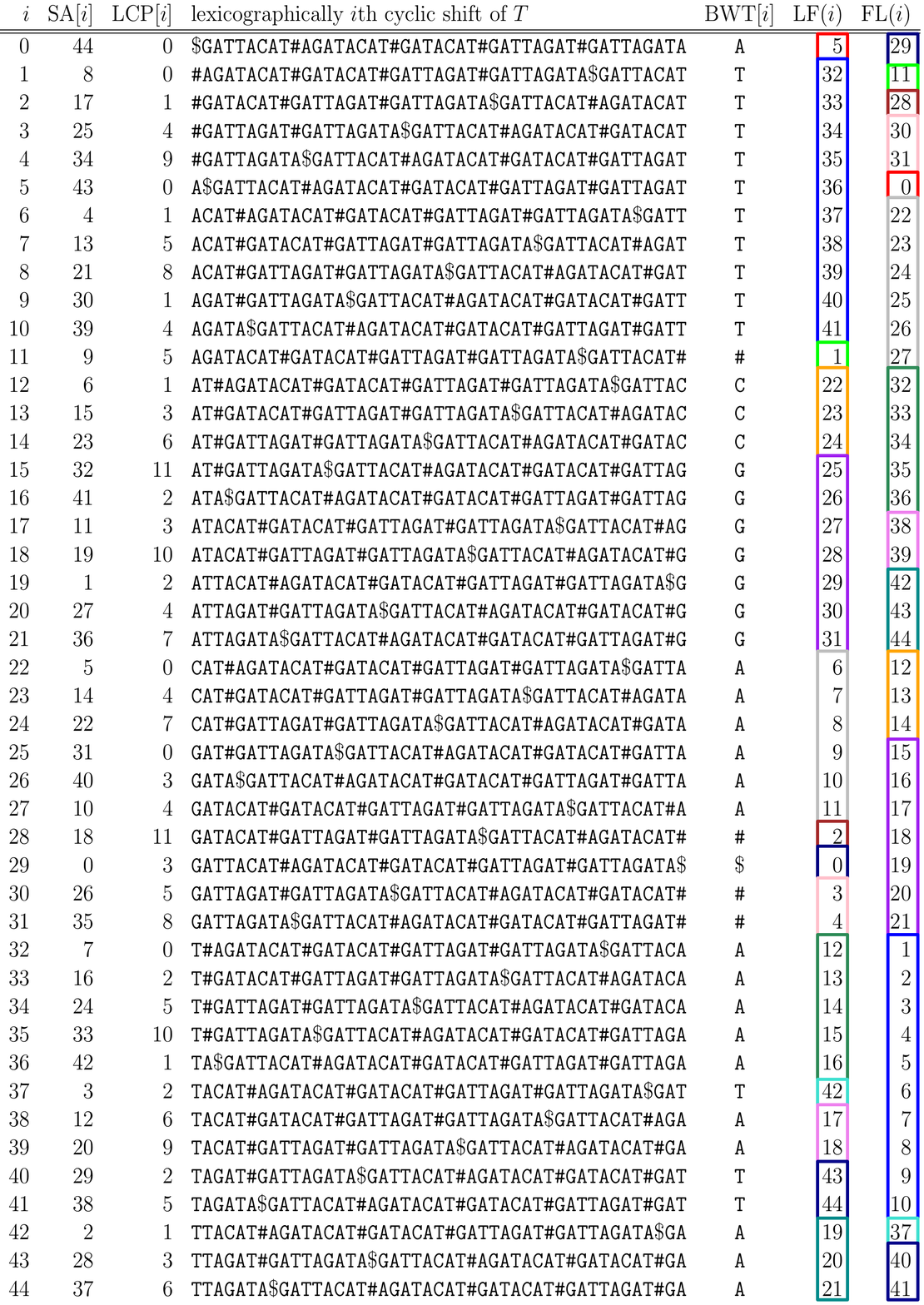}
\caption{The full table from which we conceptually start when building MONI for our example text $T = \mathtt{GATTACAT\#AGATACAT\#GATACAT\#GATTAGAT\#GATTAGATA\$}$.}
\label{tab:base_table}
\end{center}
\end{table}

For each of the value $j$ between 0 and $r - 1 = 13$, we conceptually extract from this table the starting and ending positions $\head (j)$ and $\tail (j)$ of run $j$ in the BWT, $\SA [\head (j)]$, $\SA [\tail (j)]$, $\BWT [\head (j)]$, $\LF [\head (j)]$ and the rank of the predecessor of $\LF [\head (j)]$ in the set
\[\{\head [0], \ldots, \head [r - 1]\}\,.\]
In practice we can compute the values directly without building Table~\ref{tab:base_table}, using prefix-free parsing~\cite{BGKLMM19}.

We build Table~\ref{tab:compressed_table} with these values but we sort the last two columns, which we refer to as $\mu (j)$ and $\finger (j)$.  An equivalent way to define $\mu (j)$ and $\finger (j)$, illustrated in Table~\ref{tab:base_table}, is to draw boxes corresponding to the runs in the BWT, permute those boxes according to LF, and write their starting positions in order as the $\mu (j)$ values and the numbers of the runs in the BWT covering their starting positions as the $\finger (j)$ values.  Storing Table~\ref{tab:compressed_table} takes about
\[2 r \lg (n / r) + 2 r \lg n + r \lg \sigma + 2 r\]
bits, where $\sigma$ is the size of the alphabet.  (We do not actually need to store $\tail (j) = \head (j + 1) - 1$, of course, but we include it in Table~\ref{tab:compressed_table} to simplify our explanation.)  It is within a reasonable constant factor of the most space-efficient implementation and simple to build.

\begin{table}
\begin{center}
\begin{tabular}{r|rrrrcrr}
$j$ & $\head (j)$  & $\SA [\head (j)]$ & $\tail (j)$ & $\SA [\tail (j)]$ & $\BWT [\head (j)]$ & $\mu (j)$ & $\finger (j)$ \\
\hline \hline
 0 &  0 & 44 &  0 & 44 & \tt A &   0 &  0\\
 1 &  1 &  8 & 10 & 39 & \tt T &   1 &  1\\
 2 & 11 &  9 & 11 &  9 & \tt \# &  2 &  1\\
 3 & 12 &  6 & 14 & 23 & \tt C &   3 &  1\\
 4 & 15 & 32 & 21 & 36 & \tt G &   5 &  1\\
 5 & 22 &  5 & 27 & 10 & \tt A &   6 &  1\\
 6 & 28 & 18 & 28 & 18 & \tt \# & 12 &  3\\
 7 & 29 &  0 & 29 &  0 & \tt \$ & 17 &  4\\
 8 & 30 & 26 & 31 & 35 & \tt \# & 19 &  4\\
 9 & 32 &  7 & 36 & 42 & \tt A &  22 &  5\\
10 & 37 &  3 & 37 &  3 & \tt T &  25 &  5\\
11 & 38 & 12 & 39 & 20 & \tt A &  32 &  9\\
12 & 40 & 29 & 41 & 38 & \tt T &  42 & 13\\
13 & 42 &  2 & 44 & 37 & \tt A &  43 & 13\\
\end{tabular}
\end{center}
\caption{The values we extract from Table~\ref{tab:base_table}, with the last two columns sorted.}
\label{tab:compressed_table}
\end{table}

For each suffix $P [i..m - 1]$ of $P$ from shortest to longest, MONI finds the length $\ell_i$ of the longest prefix $P [i..i + \ell_i - 1]$ of $P [i..m - 1]$ that occurs in $T$, the lexicographic rank $q_i$ of a suffix of $T$ starting with $P [i..i + \ell_i - 1]$, the starting position $\SA [q_i]$ of that suffix in $T$, and the row $j_i$ of Table~\ref{tab:compressed_table} such that $\head (j_i)$ is the predecessor of $q_i$ in that column.  We note that the $(\pos, \len)$ pairs $(\SA [q_0], \ell_0), \ldots, (\SA [q_{m - 1}, \ell_{m - 1})$ are the matching statistics $\MS [0..m - 1]$ of $P$ with respect to $T$.

Suppose we know $i$, $\ell_i$, $q_i$, $\SA [q_i]$ and $j_i$, and we want to find $\ell_{i - 1}$, $q_{i - 1}$, $\SA [q_{i - 1}]$ and $j_{i - 1}$.  If $\BWT [j_i] = P [i - 1]$ then we perform an LF step, as we describe in a moment.  If $\BWT [j_i] \neq P [i - 1]$ then we find the last row $j_i'$ above row $j_i$ with $\BWT [\head (j_i')] = P [i - 1]$, and the first row $j_i''$ below row $j_i$ with $\BWT [\head (j_i'')] = P [i - 1]$.  We use LCE queries to check whether $T [\SA [q_i]..n - 1]$ has a longer common suffix with $T [\SA [\tail (j_i')]..n - 1]$ or with $T [\SA [\head (j_i'')]..n - 1]$ and, depending on that comparison, either reset
\begin{eqnarray*}
\ell_i & = & \LCE (\SA [q_i], \SA [\tail (j_i')])\\
q_i & = & \tail (j_i')\\
\SA [q_i] & = & \SA [\tail (j_i')]\\
j_i & = & j_i'
\end{eqnarray*}
or reset
\begin{eqnarray*}
\ell_i & = & \LCE (\SA [q_i], \SA [\head (j_i'')])\\
q_i & = & \head (j_i'')\\
\SA [q_i] & = & \SA [\head (j_i'')]\\
j_i & = & j_i''\,.
\end{eqnarray*}
Now $\BWT [j_i] = P [i - 1]$, so we can proceed with the LF step.

For example, suppose $P [0..11] = \mathtt{TAGATTACATTA}$, $i = 2$ and we have already found $\ell_2 = 8$ (because {\tt GATTACAT} occurs in $T$ but {\tt GATTACATT} does not), $q_2 = 29$, $\SA [q_2] = 0$ and $j_2 = 7$.  Since $\BWT [\head (7)] = \mathtt{\$} \neq P [0] = \mathtt{A}$, we find $j_2' = 5$ and $j_2'' = 9$ and compare
\[\LCE (0, \SA [\tail (5)]) = \LCE (0, 10) = 3\]
against
\[\LCE (0, \SA [\head (9)]) = \LCE (0, 7) = 0\,.\]
Since the former LCE is longer, we set $\ell_2 = 3$, $q_2 = 27$, $\SA [q_2] = 10$ and $j_2 = 5$.

To perform an LF step with Table~\ref{tab:compressed_table} when we know $i$, $\ell_i$, $q_i$, $\SA [q_i]$ and $j_i$, we first set
\begin{eqnarray*}
\ell_{i - 1} & = & \ell_i + 1\\
q_{i - 1} & = & \mu (\pi (j_i)) + q_i - \head (j_i)\\
\SA [q_{i - 1}] & = & \SA [q_i] - 1\,,
\end{eqnarray*}
where $\pi$ is the permutation on $\{0, \ldots, r - 1\}$ that stably sorts the column $\BWT [\head (j)]$.  If we keep $\BWT [\head (j)]$ in a wavelet tree then we have fast access to $\pi$.

For our example, consider $i = 2$, $\ell_2 = 3$, $q_2 = 27$, $\SA [q_2] = 10$ and $j_2 = 5$.  Since $\BWT [\head (5)]$ is the second {\tt A} in the column $\BWT [\head (j)]$ and there are 4 characters in the column lexicographically strictly less than {\tt A}, $\pi (5) = 5$ and $\mu (5) = 6$, so we set $\ell_1 = 4$, $q_1 = 6 + 27 - 22 = 11$ and $\SA [11] = 9$.  Notice $\pi$ is similar to an LF mapping for the sequence obtained by sampling one character from each run of the BWT (but in our example $\mu$ has a fixed point at 5); in fact, it permutes the coloured boxes in Table~\ref{tab:base_table} according to LF.  It follows that $\mu (\pi (j_i)) = \LF (\head (j_i))$.  Since LF maintains the relationship between elements in the same box,
\[\LF (q_i) - \LF (\head (j_i)) = q_i - \head (j_i)\,;\]
substituting and rearranging, we obtain our formula for $q_{i - 1}$.

The last thing left for us to do during an LF step is find $j_{i - 1}$.  For this, we use the $\finger (j)$ column.  By construction, $\head (\finger (\pi (j_i)))$ is the predecessor of $\LF (\head (j_i))$ in the set
\[\{\head [0], \ldots, \head [r - 1]\}\,.\]
Therefore, since $q_{i - 1} = \LF (q_i) \geq \LF (\head (j_i))$, we can find the row $j_{i - 1}$ of Table~\ref{tab:compressed_table} such that $\head (j_{i - 1})$ is the predecessor of $q_{i - 1}$ in that column, by starting an exponential search at row $\finger (\pi (j_i))$.  This takes $O (\log r)$ time in the worst case and in practice it takes constant time.  Nishimoto and Tabei showed how to guarantee it takes constant time at the cost of increasing the size of Table~\ref{tab:compressed_table} slightly.

For more formal discussions, we refer readers to previous papers on MONI~\cite{ROLGB22} and the $r$-index~\cite{GNP20,NT21,BGR22,NKT22}.

\section{LCP access}
\label{sec:LCP}

We can support $\phi$ queries with table lookup as well: for each run $\BWT [i..j]$ in the BWT, we store $\SA [i]$ and $\SA [(i - 1) \bmod n]$ as a row; we sort the rows by their first components; and we add to each row the number of the row containing the predecessor in the first column of the second component.  Abusing notation slightly, we refer to the columns of the resulting table as $\SA [\head (j)]$, $\SA [\tail (j)]$ and $\finger (j)$.  Table~\ref{tab:phi} is for our running example, augmented with a column $\LCP [\head (j)]$ that stores the length of the longest common prefix of $T [\SA [\head (j)]..n - 1]$ and $T [\SA [\tail (j)]..n - 1]$.  Since we are storing the row containing the predecessor of each entry in $\SA [\tail (j)]$ in the column $\SA [\head (j)]$, we can encode each entry in $\SA [\tail (j)]$ as the difference between it and its predecessor in $\SA [\head (j)]$.  Analysis then shows the table then takes about $3 r \lg (n / r) + r \lg r$ bits.

\begin{table}[t]
\begin{center}
\begin{tabular}{r|rrrr}
$j$ & $\SA [\head (j)]$ & $\SA [\tail (j)]$ & $\LCP [\head (j)]$ & $\finger (j)$\\
\hline \hline
 0 &  0 & 18 &  3 &  9\\
 1 &  2 & 38 &  1 & 12\\
 2 &  3 & 42 &  2 & 12\\
 3 &  5 & 36 &  0 & 12\\
 4 &  6 &  9 &  1 &  7\\
 5 &  7 & 35 &  0 & 12\\
 6 &  8 & 44 &  0 & 13\\
 7 &  9 & 39 &  5 & 12\\
 8 & 12 &  3 &  6 &  2\\
 9 & 18 & 10 & 11 &  7\\
10 & 26 &  0 &  5 &  0\\
11 & 29 & 20 &  2 &  9\\
12 & 32 & 23 & 11 &  9\\
13 & 44 & 37 &  0 & 12
\end{tabular}
\caption{The table we use for $\phi$ queries and access to the LCP.}
\label{tab:phi}
\end{center}
\end{table}

To see how we use Table~\ref{tab:phi} to answer $\phi$ queries, suppose we know that the predecessor of 24 in $\SA [\head (j)]$ is in row 9.  Then we have
\[\phi (24)
= \SA [\tail (9)] + 24 - \SA [\head (9)]
= 10 + 24 - 18
= 16\,.\]
We know that the predecessor of 10 in $\SA [\head (j)]$ is in row $\finger (9) = 7$, but the predecessor of 16 could be in a later row.  Again, we perform an exponential search starting in row $\finger (9) = 7$ and find the predecessor 12 of 16 in row 8.  Then we have
\[\phi (16)
= \SA [\tail (8)] + 16 - \SA [\head (8)]
= 3 + 16 - 12
= 7\,.\]
Looking at rows 32 to 34 in Table~\ref{tab:base_table}, we see that indeed $\phi (24) = 16$ and $\phi (16) = 7$.  This works because, similar to the equation for LF, if $\BWT [j - 1] = \BWT [j]$ then $\phi (\SA [j] - 1) = \phi (\SA [j]) - 1$.  Again, for more formal discussions, we refer readers to previous papers on the $r$-index~\cite{GNP20,NT21,BGR22,NKT22}.

We do not know how to support random access to the LCP array quickly in $O (r \log n)$ bits, but we can use Table~\ref{tab:phi} to provide a kind of sequential access to it.  Specifically, as we use $\phi$ to enumerate the values in the SA --- without necessarily knowing the positions of the cells of the SA those values appear in --- we can use similar computations to enumerate the corresponding values in the LCP array.  In our example, since the predecessor 18 of 24 in $\SA [\head (j)]$ is in row 9, we can compute the LCP value corresponding to the SA value 24 as
\[\LCP [\SA^{-1} [24]]
= \LCP [\head (9)] + \SA [\head (9)] - 24
= 11 + 18 - 24
= 5\,.\]
Checking this, we see that $\LCP [\SA^{-1} [24]] = \LCP [34] = 5$.  Since the predecessor 12 of $\phi (24) = 16$ in $\SA [\head (j)]$ is in row 8 of Table~\ref{tab:phi},
\[\LCP [\SA^{-1} [16]]
= \LCP [\head (8)] + \SA [\head (8)] - 16
= 6 + 12 - 16
= 2\,.\]
Checking this, we see that $\LCP [\SA^{-1} [16]] = \LCP [33] = 2$.

Notice we do not use the row numbers 34 and 33 to compute the LCP value, as the SA value 24 is sufficient.  We could avoid using the inverse suffix array $\SA^{-1}$ in our formula by writing $\LCP [\SA^{-1} [24]]$ as $\PLCP [24]$, for example, where $\PLCP [0..n - 1]$ denotes the permuted LCP array~\cite{KMP09} of $T$.  The kind of sequential access we obtain to the LCP is actually random access to the PLCP array, and it is easier to explain why it works from that perspective --- because if $\BWT [j - 1] = \BWT [j]$ then $\PLCP [\SA [j] - 1] = \PLCP [\SA [j]] + 1$.\footnote{The formula for $\PLCP$ has a $+1$ where the formula for $\phi$ has a $-1$,
\begin{eqnarray*}
\phi (\SA [j] - 1) & = & \phi (\SA [j]) - 1\\
\PLCP [\SA [j] - 1] & = & \PLCP [\SA [j]] + 1\,,
\end{eqnarray*}
because if $\BWT [j - 1] = \BWT [j]$ then moving from $j$ to $LF (j)$ decrements the SA entry but increments the LCP entry.}  Nevertheless, we present our results in terms of the LCP and $\SA^{-1}$ because we will use them later in conjunction with $\phi$ queries to enumerate the values in LCP intervals.

Symmetric to using Table~\ref{tab:phi} to support $\phi$ queries, we can use a table to support $\phi^{-1}$ queries.  In fact, the $(\SA [\head (j)], \SA [\tail (j)])$ pairs in the table are the same, but sorted by their second components; now we add to each row the number of the row containing the predecessor in the second column of the first component.  Since we are storing the row containing the predecessor of each entry in $\SA [\head (j)]$ in the column $\SA [\tail (j)]$, we can encode each entry in $\SA [\head (j)]$ as the difference between it and its predecessor in $\SA [\tail (j)]$.  Analysis then shows the table takes about $2 r \lg (n / r) + r \lg r$ bits.  Table~\ref{tab:phiinv} is for supporting $\phi^{-1}$ queries on our running example.  For example, if we know that the predecessor of 7 in $\SA [\tail (j)]$ is in row 1, then we can compute
\[\phi^{-1} (7)
= \SA [\head (1)] + 7 - \SA [\tail (1)]
= 12 + 7 - 3
= 16\]
and we can find the row containing the predecessor of 16 in $\SA [\tail (j)]$ with an exponential search starting at row $\finger (1) = 3$ (and ending in the same row).  We can then compute
\[\phi^{-1} (16)
= \SA [\head (3)] + 16 - \SA [\tail (3)]
= 18 + 16 - 10
= 24\]
and we can find the row 6 containing the predecessor of 24 in $\SA [\tail (j)]$ with an exponential search starting at row $\finger (3) = 4$.

\begin{table}[t]
\begin{center}
\begin{tabular}{r|rrr}
$j$ & $\SA [\head (j)]$ & $\SA [\tail (j)]$ & $\finger (j)$\\
\hline \hline
 0 & 26 &  0 &  6\\
 1 & 12 &  3 &  3\\
 2 &  6 &  9 &  1\\
 3 & 18 & 10 &  4\\
 4 &  0 & 18 &  0\\
 5 & 29 & 20 &  6\\
 6 & 32 & 23 &  6\\
 7 &  7 & 35 & 11\\
 8 &  5 & 36 &  1\\
 9 & 44 & 37 & 13\\
10 &  2 & 38 &  0\\ 
11 &  9 & 39 &  2\\
12 &  3 & 42 &  1\\
13 &  8 & 44 &  1
\end{tabular}
\caption{The table we use for $\phi^{-1}$ queries.}
\label{tab:phiinv}
\end{center}
\end{table}

With these two $O (r \log n)$-bit tables, given $k$, $j$ and $\SA [j]$, we can compute $\SA [j - k + 1..j + k - 1]$ and $\LCP [j - k + 1..j + k - 1]$ in $O (k \log r) \subseteq O (k \log n)$ time.  (Actually, we can achieve that bound even without the $\finger (j)$ columns in the tables, but Brown et al.'s results suggest those will provide a significant speedup in practice.)  With Nishimoto and Tabei's modification, we can reduce that to $O (k)$ time while keeping the tables in $O (r \log n)$ bits; this would slightly improve the time bound we give in the next section to $O (m (k + \log n))$.

\begin{lemma}
\label{lem:phi}
We can store two $O (r \log n)$-bit tables such that, given $k$, $j$ and $\SA [j]$, we can compute $\SA [j - k + 1..j + k - 1]$ and $\LCP [j - k + 1..j + k - 1]$ in $O (k \log n)$ time. 
\end{lemma}

\section{Finding $k$-MEMs with Lemma~\ref{lem:phi}}
\label{sec:slow}

We store the tables described in Sections~\ref{sec:MONI} and~\ref{sec:LCP} for $T$, which add $O (r \log n)$ bits to MONI.  Given $P$ and $k$, we find the MEMs of $P$ with respect to $T$ as before but then, from each $\SA [q_i]$, we use Lemma~\ref{lem:phi} to find $\LCP [q_i - k + 2..q_i + k - 1]$ in $O (k \log n)$ time.

For example, suppose that $P [0..11] = {\tt TAGATTACATTA}$, as in Section~\ref{sec:MONI}, and $k = 3$.  Starting with $q_{12} = 22$, with MONI we compute the values shown in columns $q_i$, $\SA [q_i]$, $\ell_i$ and $\BWT [q_i]$ of Table~\ref{tab:MONI_run}.    (It is important that we choose $q_i$ to be the endpoint of a run, since we store SA entries only at those positions, but this is true also for MONI.)  The crossed out values are the ones we replace because $\BWT [q_i] \neq P [i]$.  If we look at the original $\SA [q_i]$ and $\ell_i$ values, before any replacements, we obtain the matching statistics
\[\MS [0..11] = (38, 5), (9, 4), (0, 8), (1, 7), (2, 6), (39, 5), (21, 4), (22, 3), (1, 4), (2, 3), (3, 2), (4, 1)\]
of $P$ with respect to $T$, with $(\pos, \len)$ pair $\MS [i]$ indicating the starting position $\MS [i].\pos$ in $T$ of an occurrence of the longest prefix of $P [i..m - 1]$ that occurs in $T$, and the length $\MS [i].\len$ of that prefix.

From the matching statistics, it is easy to compute the MEMs $P [0..4] = \mathtt{TAGAT}$, $P [3..9] = \mathtt{ATTACAT}$ and $P [8..11] = \mathtt{ATTA}$ of $P$ with respect to $T$: a MEM starts at any position $i$ such that $i = 0$ or  $\MS [i - 1].\len \leq \MS [i].\len$.  For each $i$, after we compute $q_i$, $\SA [q_i]$ and $\ell_i$ (and before we replace them, if we do), we use Lemma~\ref{lem:phi} to compute the sub-intervals of length 4 in column $\LCP [q_i - 1..q_i + 2]$.

\begin{table}[t]
\begin{center}
\begin{tabular}{r|crrrr|crr}
$i$ & $P [i]$ & $q_i$ & $\SA [q_i]$ & $\ell_i$ & $\BWT [q_i]$ & $\LCP [q_i - 1..q_i + 2]$ & $L_i$ & $\min (\ell_i, L_i)$\\
\hline
12 &         & 22\, &  5\, & 0\, & \tt A\, \\
11 & {\tt A} &  6\, &  4\, & 1\, & \tt T\, & [0, 1, 5, 8] & 5 & 1\\
10 & {\tt T} & 37\, &  3\, & 2\, & \tt T\, & [1, 2, 6, 9] & 6 & 2\\
 9 & {\tt T} & 42\, &  2\, & 3\, & \tt A\, & [5, 1, 3, 6] & 3 & 3\\
 8 & {\tt A} & 14 \sout{\,19\,} & 23 \sout{\,1\,} & 2 \sout{\,4\,} & \tt C \sout{\,G\,} & [10, 2, 4, 7] & 4 & 4\\
 7 & {\tt C} & 24\, & 22\, & 3\, & \tt A\, & [4, 7, 0, 3] & 4 & 3\\
 6 & {\tt A} &  8\, & 21\, & 4\, & \tt T\, & [5, 8, 1, 4] & 5 & 4\\
 5 & {\tt T} & 37 \sout{\,39\,} & 3 \sout{\,20\,} & 5\, & \tt T \sout{\,A\,} & [6, 9, 2, 5] & 6 & 5\\
 4 & {\tt T} & 42\, &  2\, & 6\, & \tt A\, & [5, 1, 3, 6] & 3 & 3\\
 3 & {\tt A} & 19\, &  1\, & 7\, & \tt G\, & [10, 2, 4, 7] & 4 & 4\\
 2 & {\tt G} & 27 \sout{\,29\,} & 10 \sout{\,0\,} & 3 \sout{\,8\,} & \tt A \sout{\,\$\,} & [11, 3, 5, 8] & 5 & 5\\
 1 & {\tt A} & 10 \sout{\,11\,} & 39 \sout{\,9\,} & 4\, & \tt T \sout{\,\#\,} & [4, 5, 1, 3] & 4 & 4\\
 0 & {\tt T} & 41\, & 38\, & 5\, & \tt T\, & [2, 5, 1, 3] & 2 & 2
\end{tabular}
\caption{With MONI we compute the values shown in columns $q_i$, $\SA [q_i]$, $\ell_i$ and $\BWT [q_i]$ on the left side of the table, and from those we can compute the matching statistics and MEMs of $P [0..11] = \mathtt{TAGATTACATTA}$ with respect to $T [0..44] = \mathtt{GATTACAT\#AGATACAT\#GATACAT\#GATTAGAT\#GATTAGATA\$}$.  After we have computed the values on the left side of the table, we can also compute the values in columns $\LCP [q_i - k + 2..q_i + k - 1]$, $L_i$ and $\min (\ell_i, L_i)$ on the right side of the table, and from those we can compute the $3$-MEMs of $P$ with respect to $T$.}
\label{tab:MONI_run}
\end{center}
\end{table}

We scan each interval $\LCP [q_i - k + 2..q_i + k - 1]$ in $O (k)$ time and find a sub-interval of length $k - 1$ such that the minimum LCP value $L_i$ in that sub-interval is maximized.  This LCP sub-interval corresponds to an sub-interval of length $k$ in $\SA [q_i - k + 1..q_i + k - 1]$ containing the starting positions of $k$ suffixes of $T$ --- including $T [\SA [q_i]..n - 1]$ itself --- whose common prefix with $T [\SA [q_i]..n - 1]$ has the maximum possible length $L_i$.

In our example, we scan each interval in column $\LCP [q_i - 1..q_i + 2]$ of Table~\ref{tab:MONI_run} and find the sub-interval of length 2 such that the minimum LCP value $L_i$ is maximized.  If we check Table~\ref{tab:base_table}, we find that the longest prefix of $T [4..44] = \mathtt{ACAT\#AGATA}\ldots$ that occurs at least 3 times in $T$ indeed has length $L_{11} = 5$, the longest prefix of $T [3..44] = \mathtt{TACAT\#AGATA}\ldots$ that occurs at least 3 times in $T$ has indeed has length $L_{10} = 6$, the longest prefix of $T [2..44] = \mathtt{TTACAT\#AGATA}\ldots$ indeed has length $\ell_9' = 3$, and so on.

Since the common prefix of $P [i..m - 1]$ and $[\SA [q_i]$ has the maximum possible length $\ell_i$, the longest prefix of $P [i..m - 1]$ that occurs at least $k$ times in $T$ has length $\min (\ell_i, L_i)$.  Computing $\min (\ell_i, L_i)$ for each $i$ takes a total of $O (k m \log n)$ time.  The values $\min (\ell_i, L_i)$ are something like a parameterized version of the lengths in the matching statistics: $\min (\ell_i, L_i)$ is the length of the longest prefix of $P [i..m - 1]$ that occurs at least $k$ times in $T$.

We can compute the $k$-MEMs of $P$ with respect to $T$ from the $\min (\ell_i, L_i)$ values the same way we compute MEMs from the lengths in the matching statistics: a $k$-MEM starts at any position $i$ such that $i = 0$ or $\min (\ell_{i - 1}, L_{i - 1}) \leq \min (\ell_i, L_i)$.  In our example,
\begin{eqnarray*}
\min (\ell_0, \ell_0') & \leq & \min (\ell_1, \ell_1')\ =\ 4\\
\min (\ell_1, \ell_1') & \leq & \min (\ell_2, \ell_2')\ =\ 5\\
\min (\ell_4, \ell_4') & \leq & \min (\ell_5, \ell_5')\ =\ 5\\
\min (\ell_7, \ell_7') & \leq & \min (\ell_8, \ell_8')\ =\ 4
\end{eqnarray*}
and so the $k$-MEMs are $P [0..1] = \mathtt{TA}$, $P [1..4] = \mathtt{AGAT}$, $P [2..6] = \mathtt{GATTA}$, $P [5..9] = \mathtt{TACAT}$ and $P [8..11] = \mathtt{ATTA}$.

We can compute $\min (\ell_i, L_i)$ as soon as we have we have computed $\SA [q_i]$ and $\ell_i$, so we can compute the $k$-MEMs of $P$ with respect to $T$ online.

\begin{theorem}
\label{thm:query_time}
Suppose we have MONI for a text $T [0..n - 1]$ whose BWT consists of $r$ runs.  We can add $O (r \log n)$ bits to MONI such that, given $P [0..m - 1]$ and $k$, we can find the $k$-MEMs of $P$ with respect to $T$ online in $O (k \log n)$ time per character of $P$.
\end{theorem}

\section{Finding $k$-MEMs with precomputed values}
\label{sec:fast}

Suppose the interval of length $k$ that we find in SA for $P [i..m - 1]$, following the procedures in Section~\ref{sec:slow}, is $\SA [s_i..s_i + k - 1]$ and $\BWT [s_i] = \cdots = \BWT [s_i + k - 1] = P [i - 1]$.  Then $\min (\ell_{i - 1}, L_{i - 1}) = \min (\ell_i, L_i) + 1$ and we can find the interval for $P [i - 1..m - 1]$ with an LF queries for $s_i$, in $O (\log n)$ time.  This means we need the results of Section~\ref{sec:LCP} only when at least one character in $\BWT [s_i..s_i + k - 1]$ is not equal to $P [i - 1]$.

First, suppose $\BWT [q_i] \neq P [i - 1]$.  Following the procedures in Section~\ref{sec:MONI}, MONI resets $q_i$ to the endpoint $b$ of a run in the BWT, reset $\ell_i$, and then compute $q_{i - 1} = \LF (b)$.  Following the procedures in Section~\ref{sec:slow}, we compute $\LCP [q_{i - 1} - k + 2..q_{i - 1} + k - 1]$ and scan it to compute the interval $\SA [s_{i - 1}..s_{i - 1} + k - 1]$ for $P [i - 1..m - 1]$.

If we are given $k$ at construction time, however, then for every endpoint $b$ of a run in the BWT, we can precompute
\begin{itemize}
\item the sub-interval of length $k - 1$ of $\LCP [\LF (b) - k + 2..\LF (b) + k - 1]$ that maximizes the minimum value $L (b)$ in the sub-interval,
\item that value $L (b)$.
\end{itemize}
With this information, we do not need the results of Section~\ref{sec:LCP} for this case either, and can handle it in $O (\log n)$ time as well.  Since the sub-interval we store for $b$ starts between $\LF (b) - k + 2$ and $\LF (b) + k - 1$, we can store it in $O (\log k)$ bits as an offset.  This means we store $O (r \log k)$ bits on top of at most $2 r$ LCP values, or $O (r \log n)$ bits in total.

The remaining case is when $\BWT [q_i] = P [i - 1]$ but some of the other characters in $\BWT [s..s + k - 1]$ are not equal to $P [i - 1]$.  If $\BWT [q_i]$ is the end of a run, then we can proceed as in the previous case in $O (\log n)$ time, using our precomputed values for $q_i$ (but without resetting $q_i$ and $\ell_i$).  Otherwise, we claim we can choose such a character $\BWT [b]$, set
\[\ell_i = \min (\LCE (\SA [q_i], \SA (b)), \ell_i)\]
and $q_i = b$, and then proceed as in the previous case in $O (\log n)$ time, and still be sure of obtaining the correct $k$-MEMs of $P$ with respect to $T$.  (Continuing to run MONI with the new values of $q_i$ and $\ell_i$ may not give us the correct MEMs, however.)  To be able to change $q_i$ and $\ell_i$ this way, it is important that we now work online, instead of running MONI on $P$ and then using the results to find the $k$-MEMs.

To see why our claim holds, assume our query has worked correctly so far, so
\[T [\SA [q_i]..\SA [q_i] + \min (\ell_i, L_i) - 1]
= T [\SA [b]..\SA [b] + \min (\ell_i, L_i) - 1]\]
is the longest prefix of $P [i..m - 1]$ that occurs at least $k$ times in $T$.  Therefore, the $k$-MEMs starting in $P [0..i - 1]$ are all completely contained in $P [0..i + \min (\ell_i, L_i) - 1]$.  It follows that resetting
\[\ell_i = \min (\LCE (\SA [q_i], \SA (b)), \ell_i)\]
and $q_i = b$ does not affect the set of $k$-MEMs we find that start in $P [0..i - 1]$.

Figure~\ref{fig:pseudo-code} shows pseudo-code for how we find $k$-MEMs with precomputed values.  Table~\ref{tab:colourful_table} shows the offsets and $L (b)$ values for our example, surrounded by coloured boxes on the right, with each offset indicating how far above $\LF (b)$ the sub-interval starts.  The coloured boxes on the left indicate the sub-interval itself and the longest common prefix of the suffixes starting in the sub-interval of the SA.

\begin{figure}[t]
\begin{center}
{\tt \begin{tabular}{l}
if $\BWT [s_i] = \cdots = \BWT [s_i + k - 1]= P [i - 1]$ then\\
\ \ $q_{i - 1} \leftarrow \LF (q_i)$\\
\ \ $\ell_{i - 1} \leftarrow \ell_i + 1$\\
\ \ $L_{i - 1} \leftarrow L_i + 1$\\
\ \ $s_{i - 1} = \LF (s_i)$\\
else\\
\ \ if $\BWT [q_i] \neq P [i - 1]$ then\\
\ \ \ \ reset $q_i$ and $\ell_i$ as {\rm MONI} does\\
\ \ else if $\BWT [q_i - 1]$ is not the end of a run\\
\ \ \ \ choose $b$ in $[s_i..s_i + k - 1]$ with $\BWT [b] = P [i - 1]$ at the end of a run\\
\ \ \ \ $\ell_i \leftarrow \min (\LCE (\SA [q_i], \SA [b]), \ell_i)$\\
\ \ \ \ $q_i \leftarrow b$\\
\ \ end if\\
\ \ $q_{i - 1} \leftarrow \LF (q_i)$\\
\ \ $\ell_{i - 1} \leftarrow \ell_i + 1$\\
\ \ $L_{i - 1} \leftarrow L (q_i)$\\
\ \ $s_{i - 1} \leftarrow \LF (q_i) - \offset (q_i)$\\
end if
\end{tabular}}
\caption{Pseudo-code for how we find $k$-MEMs with precomputed values.}
\label{fig:pseudo-code}
\end{center}
\end{figure}

\begin{table}
\begin{center}
\includegraphics[width=\textwidth]{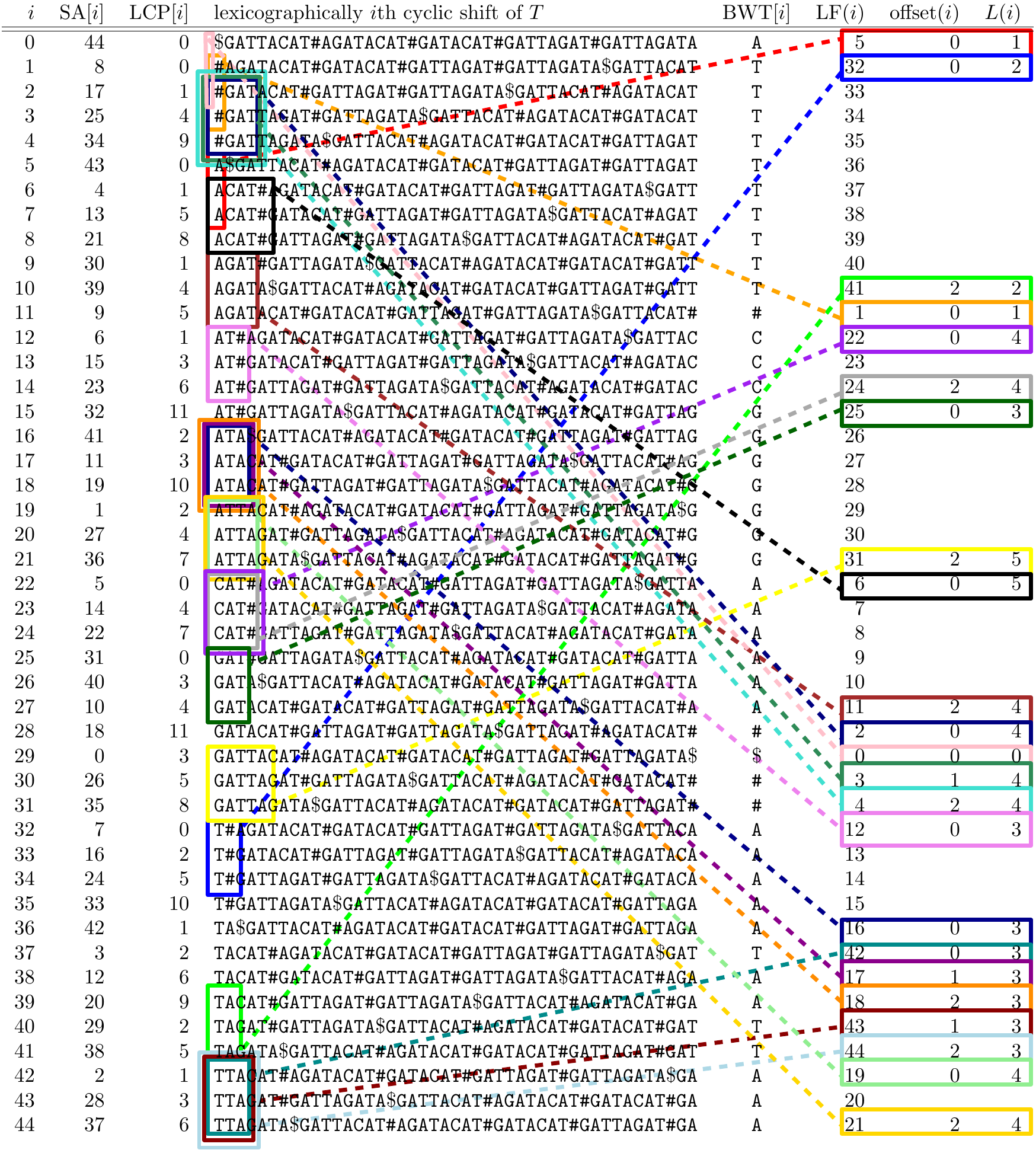}
\caption{The table showing the precomputed values we use to find 3-MEMs with respect to our example $T = \mathtt{GATTACAT\#AGATACAT\#GATACAT\#GATTAGAT\#GATTAGATA\$}$.}
\label{tab:colourful_table}
\end{center}
\end{table}

For our example, suppose we again start with $q_{12} = 22$ and $\ell_{12} = 0$.  Since $\BWT [q_{12}] = P [11] = \mathtt{A}$, we set $q_{11} = \LF (22) = 6$ and $\ell_{11} = \ell_{12} + 1 = 1$.  The values $\offset (22) = 0$ and $L (22) = 5$ in the black rectangle in Table~\ref{tab:colourful_table} tell us to set $s_{11} = \LF (22) - 0 = 6$ and $L_{11} = 5$.  This means the suffixes of $T$ with starting points in
\[\SA [6..8] = [17, 25, 34]\]
have a longest common prefix of length 5, which starts with the longest prefix of $P [11]$ that occurs at least 3 times in $T$.  This longest prefix has length $\min (\ell_{11}, L_{11}) = 1$ --- so it is just $P [11] = \mathtt{A}$.  After this initial setup, we can fill in Table~\ref{tab:precomputed} according to the pseudo-code in Figure~\ref{fig:pseudo-code}, with crossed out values again indicating those that are replaced.

\begin{table}
\begin{center}
\begin{tabular}{rrrrcrcrc}
$i$ & $q_i$ & $\ell_i$ & $L_i$ & $\min (\ell_i, L_i)$ & $s_i$ & $P [i - 1]$ & $\BWT [q_i]$ & $\BWT [s_i..s_i + 2]$\\
\hline
12 & 22 & 0 &   &   &    & \tt A & \tt A\\
11 &  6 & 1 & 5 & 1 &  6 & \tt T & \tt T & \tt TTT\\
10 & 37 & 2 & 6 & 2 & 37 & \tt T & \tt T & \tt TAA\\
 9 & 42 & 3 & 3 & 3 & 42 & \tt A & \tt A & \tt AAA\\
 8 & 14 \sout{\,19\,} & 2 \sout{\,4\,} & 4 & 4 & 19 & \tt C & \tt C \sout{\,G\,} & \tt GGG\\
 7 & 24 & 3 & 4 & 3 & 22 & \tt A & \tt A & \tt AAA\\
 6 &  8 & 4 & 5 & 4 &  6 & \tt T & \tt T & \tt TTT\\
 5 & 37 \sout{\,39\,} & 5 & 6 & 5 & 37 & \tt T & \tt T \sout{\,A\,} & \tt TAA\\
 4 & 42 & 6 & 3 & 3 & 42 & \tt A & \tt A & \tt AAA\\ 
 3 & 19 & 7 & 4 & 4 & 19 & \tt G & \tt G & \tt GGG\\
 2 & 27 \sout{\,29\,} & 3 \sout{\,8\,} & 5 & 5 & 29 & \tt A & \tt A \sout{\,\$\,} & \tt \$\#\#\\
 1 & 10 \sout{\,11\,} & 4 & 4 & 4 &  9 & \tt T & \tt T \sout{\,\#\,} & \tt TT\#\\
 0 & 41 & 5 & 2 & 2 & 39 &      & \tt T & \tt ATT
\end{tabular}
\caption{The values we compute (except $\BWT [s_i..s_i + 2]$, which we include here only for clarity) while finding the 3-MEMs of $P [0..11] = \mathtt{TAGATTACATTA}$ with respect to our example $T = \mathtt{GATTACAT\#AGATACAT\#GATACAT\#GATTAGAT\#GATTAGATA\$}$.}
\label{tab:precomputed}
\end{center}
\end{table}

\begin{theorem}
\label{thm:construction_time}
Suppose we have MONI for a text $T [0..n - 1]$ whose BWT consists of $r$ runs.  Given $k$, we can add $O (r \log n)$ bits to MONI such that, given $P [0..m - 1]$, we can find the $k$-MEMs of $P$ with respect to $T$ online in $O (\log n)$ time per character of $P$.
\end{theorem}

\section{Conclusion}

We have shown, first, how we can add $O (r \log n)$ bits to MONI for a text $T [0..n - 1]$, where $r$ is the number of runs in the BWT of $T$, such that if we are given $k$ at query time with $P [0..m - 1]$, then we can find the $k$-MEMs of $P$ with respect to $T$ online in $O (k \log n)$ time per character of $P$.  We have then shown how, if we are given $k$ at construction time, we can add $O (r \log k)$ bits and at most $2 r$ LCP values --- which are $O (r \log n)$ bits in total --- such that we can find the $k$-MEMs of $P$ with respect to $T$ online in $O (\log n)$ time per character of $P$.  Along the way, we have also shown how to extend $\phi$ queries to support sequential access to the LCP, which may be of independent interest.

Although we have not discussed construction, we expect it will not be difficult to modify prefix-free parsing~\cite{BGIKLMNPR21} to build our tables for $\phi$, LCP and $\phi^{-1}$ queries.  Once we can support those queries, we can use them to compute $k$-MEMs in $O (k m \log n)$ time, or to build in $O (k r)$ time the table of precomputed values that we need to compute $k$-MEMs in $O (m \log n)$ time.  In fact, once we have built the tables for $\phi$, LCP and $\phi^{-1}$ queries --- which take $O (r)$ space but may be significantly larger than our table of precomputed values --- then we can store them in external memory and recover them only when we want to build a table of precomputed values for a different choice of $k$.

We believe our approach is a practical extension of MONI and we are currently implementing it.  One possible application might be to index two genomic databases (possibly with two different values of $k$), one of haplotypes from people with symptoms of a genetic disease and one of haplotypes from people without; then, as the first step in a bioinformatics pipeline, we could use those indexes to mine for substrings that are common in one database and not in the other. 
 We think this application is interesting because, except for a remark in Bannai et al.'s paper about potentially applying MEM-finding to rare-disease diagnosis, the $r$-index and MONI have so far been considered only as tools for pangenomic {\em alignment}, and this is an application to pangenomic {\em analysis}.  If the disease is recessive or multifactorial then variations associated with it are likely to be present in the both databases, so MEM-finding is unlikely to detect them; those variations could be more frequent in the first database, however, so $k$-MEM-finding may still be useful.

\bibliographystyle{plain}
\bibliography{MONI-k}

\end{document}